\documentclass[%
 reprint,
 showpacs,
 bibnotes,
 amsmath,amssymb,
 aps,
 prl,
]{revtex4-1}

\usepackage{graphicx}
\usepackage{dcolumn}
\usepackage{bm}

\usepackage{graphicx}
\usepackage{subfigure}
\usepackage{upgreek}
\usepackage{natbib}
\usepackage{color}
\usepackage{ulem}

\begin{document}

\preprint{APS/123-QED}

\title{Aerodynamic Heating in Hypersonic Boundary Layers}
\affiliation{State Key Laboratory of Turbulence and Complex Systems,
Collaborative Innovation Center of Advanced Aero-Engine, Peking University, Beijing 100871, China}
\affiliation{Department of Mechanical Engineering, Virginia Commonwealth University}
\author{Yiding Zhu}
\author{Xi Chen}
\author{Jiezhi Wu}
\author{Shiyi Chen}
\author{Cunbiao Lee}
\altaffiliation[ ]{cblee@mech.pku.edu.cn}
\affiliation{State Key Laboratory of Turbulence and Complex Systems,
Collaborative Innovation Center of Advanced Aero-Engine,\\Peking University, Beijing 100871, China}
\author{Mohamed Gad-el-Hak} \affiliation{Department of Mechanical \& Nuclear Engineering, Virginia Commonwealth University, Richmond, VA 23284, USA}
\date{\today}

\begin{abstract}

The evolution of multi-mode instabilities in a hypersonic boundary layer and their effects on aerodynamic heating are investigated.  Experiments are conducted in a Mach 6 wind tunnel using Rayleigh-scattering flow visualization, fast-response pressure sensors, fluorescent temperature-sensitive paint (TSP), and particle image velocimetry (PIV).  Calculations are also performed based on both parabolized stability equations (PSE) and direct numerical simulations (DNS). It is found that second-mode dilatational waves, accompanied by  high-frequency alternating fluid compression and expansion, produce intense aerodynamic heating in a small region that rapidly heats the fluid passing through it.  As a result, the surface temperature rapidly increases  and results in an overshoot over the nominal transitional value. When the dilatation waves decay downstream, the surface temperature decreases gradually until transition is completed. A theoretical analysis is  provided to interpret the temperature distribution affected by the aerodynamic heating.

\begin{description}

\item[PACS numbers]
 47.20.Ft, 47.40.Ki

\end{description}
\end{abstract}

\pacs{Valid PACS appear here}
\maketitle


The tragedy of the space shuttle Columbia thirteen years ago is still a reminder of  how accurate knowledge about where and why, and with what strength, strong heating peaks occur at the wall is of crucial importance to the performance and safety of hypersonic vehicles.  A local damage to the thermal protection system (TPS) might lead to a disastrous accident \cite{Report2003}. Under  real flight condition, laminar-to-turbulence transition is one of the most important sources to bring in uncertain aerodynamic heating that might  adversely impact  the vehicle's TPS \cite{Berry2006,Greene2006,McGinley2006}.   It is commonly acknowledged that  enhanced heat transfer is related to the evolution of instability waves, which in turn modifies skin friction. Extensive studies have been performed using physical and numerical experiments as well as theoretical analyses \cite{Hopkins1972,Brostmeyer1984,Horvath2002,Wadhams2008,Berridge2010,Franko2013,Sivasubramanian2015}.

A new mechanism in the transition process, absent in low Mach-number flows, is the instability of dilatational (longitudinal) waves  \cite{Emanuel1992,Gad-el-Hak1995}, such as second- and higher-instability modes \cite{Mack1969,Stetson1992,Fedorov2011,Zhang2013,Zhang2015,Zhu2016}.   Those play increasingly more important role as Mach number increases, as demonstrated in our previos studies \cite{Zhang2015,Zhu2016}. Recent direct numerical simulations  \cite{Franko2013,Sivasubramanian2015} and experiments \cite{Berridge2010} have observed that  the second mode dominates the transition process, and additionally that the wall heat transfer exhibits a local `overshoot' peak close to the end of the transitional region. However, no physical explanation has been given in terms of fundamental thermo-aerodynamics.

\begin{figure}
\includegraphics[width=8cm]{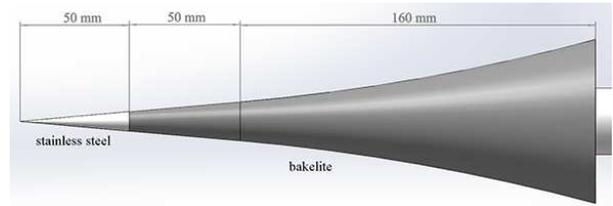}
\caption{\label{fig:cone} Schematic of the model. }
\end{figure}

\begin{figure*}
\includegraphics[width=16 cm]{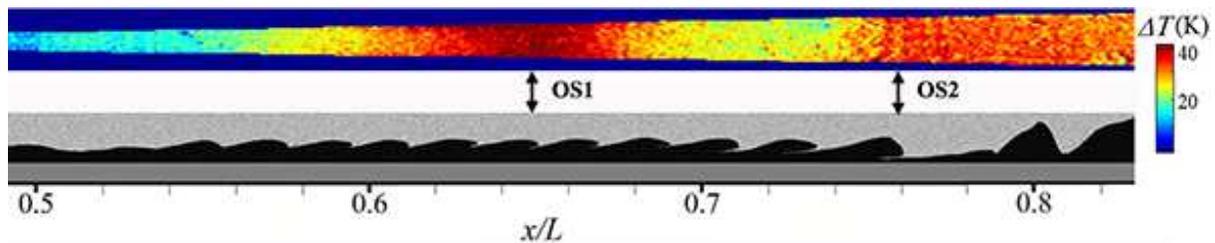}
\caption{\label{fig:tv} Comparison of  surface temperature distribution with change in flow structures during laminar-to-turbulence transition. Top:\ surface temperature rise distribution measured by TSP; bottom:\ flow visualization results using Rayleigh-scattering technique. Arrows OS1 and OS2 indicate the two regions where temperature exhibits strong overshoots.}
\end{figure*}

In this Letter, a combined experimental, numerical, and theoretical study of instability modes and their relevance to surface aerodynamic heating is reported. The evolution of instabilities and wall-temperature distributions are  observed via physical and numerical experiments. The present study not only confirms the previously observed heat-transfer overshoots in the second-mode-dominated transition process, but also reveals  a new physical mechanism to promptly generate this heating impact and its strong connection to instability evolution. For the first time a direct relation is discovered  between the local `overshoot' heating peak before transition and second-mode-induced dilatation viscous dissipation. Such relation has not been  clearly addressed in previous work.

The present experiments are carried out in the Mach 6 quiet wind tunnel (M6QT) at Peking University.  The  tunnel can run at either noisy or quiet flow condition with the nozzle throat suction valve, respectively, closed or open. Owing to the upper limit of unit Reynolds number for quiet flow condition, natural transition does not occur up to the end of the model. Therefore, the wind tunnel runs at the same noisy condition in the present work as in our previous one \cite{Zhu2016}.

The freestream stagnation temperature and pressure are, respectively, 430~K and 0.9~MPa. The freestream velocity, Mach number, and unit Reynolds number are, respectively,  $870$ m/s, 6, and $9.0\times10^{6}$ per meter. The model used is a flared, bakelite-made cone, which is installed along the centerline of the nozzle with zero angle of attack (Fig.~1). The tip of the model has a nominal 50~$\mu$m radius. The origin of the coordinate system is located at the cone's tip, with $x$ being the streamwise coordinate along the cone's surface, $y$ the coordinate normal to that surface, and $z$ the transverse coordinate normal to the $x$--$y$ plane.

Surface  pressure is measured using flush-mounted PCB 132A31 pressure transducers, located at 13 ports along an axial ray on the model from $x$=100~mm to 240~mm, with a spacing of 10~mm. The flow field is visualized using Rayleigh-scattering technique, and investigated quantitatively using a modified PIV method. The usage of PCB sensors, Rayleigh-scattering visualization, and PIV is described in detail in our prior work \cite{Zhang2015,Zhu2016}.   To investigate the surface temperature distribution, fluorescent TSP  \cite{Liu2005} is painted along another ray, from $x$=100 mm to 240 mm, 30$^\circ$ azimuthally away from the PCB ray.   The relation between the intensity ratios of TSP and temperatures at different streamwise positions is calibrated in a thermostat.    The flow is calculated based on PSE and DNS, using the same flow parameters as in the experiments.

\begin{figure*}
\includegraphics[width=16 cm]{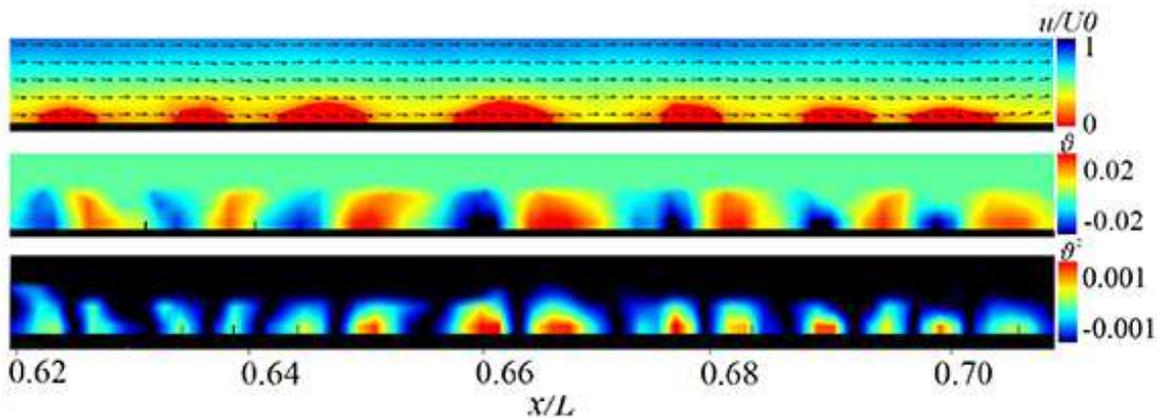}
\caption{\label{fig:PIV} PIV results of boundary layer development near OS1 region. Top:\ velocity magnitude normalized with freestream velocity; middle:\ dilatation $\vartheta =\partial_x u+\partial_y v$; bottom:\ square of dilatation $\vartheta^2$.}
\end{figure*}

Fig.~\ref{fig:tv} compares the surface temperature distribution measured using TSP with a visualization of the entire  laminar-to-turbulence transition process. The flow is from left to right. As shown, the boundary layer is laminar in the upstream region, but then regular rope-like structures appear at $x/L$=0.55, which is typical of second-mode instability waves because their wavelength is twice the boundary layer thickness. The second-mode waves grow, saturate, and finally decay to zero before $x/L$=0.75. Corresponding to the appearance of second-mode waves, the surface temperature increases initially  then experiences a fast growth and decay in region OS1 between $x/L$=0.61 and 0.68. This is followed by a second fast growth in region OS2 at $x/L$=0.74, where transition to turbulent boundary layer occurs. The physical mechanism underlying the temperature overshoot at region OS2 is relatively more understood \cite{Franko2013,Sivasubramanian2015}.  The process  is directly related to laminar-to-turbulence transition.  However, the mechanism that leads to temperature overshoot at region OS1 is not clear.  This  is the main concern of the present work. Note that heating in region OS1 is more intense than that in region OS2, and therefore region OS1 is potentially more dangerous.

Fig.~\ref{fig:PIV} presents PIV measurements of the instantaneous flow field near OS1, including the velocity magnitude normalized by the freestream velocity, the dilatation $\vartheta =\partial_x u+\partial_y v$, and $\vartheta^2$. As seen from the flow structures in Fig.~\ref{fig:tv}, the appearance of second-mode instability is associated with dilatation processes below the sonic line. For hypersonic flow, the freestream is supersonic relative to the disturbances in the boundary layer so that these disturbances will introduce Mach waves that reflect between the sonic line and solid wall. The Mach-wave series in turn generates  strong expansion and compression in the narrow subsonic near-wall regime, which inevitably bring in dilatational viscous dissipation characterized by $\vartheta^2$, as shown in the bottom plot of Fig.~\ref{fig:PIV}.

Indeed, as is well known, the dissipation function per unit volume $\Phi$ can be explicitly split into  contribution from vorticity $\omega$ and dilatation $\vartheta$ \cite{Lele1994,wu2006}:

\begin{equation}
\Phi=\mu\omega^2+\mu^\prime\vartheta^2+ {\rm BI},
\end{equation}
where BI represents a boundary integral, and $\mu^\prime =\mu_b+4\mu/3$ is the longitudinal viscosity, with $\mu$ and $\mu_b$ being, respectively, the shear and bulk viscosities. Emanuel \cite{Emanuel1992} and Gad-el-Hak \cite{Gad-el-Hak1995} recognized the important role of $\mu_b$ in high-Mach-number flows. The dilatation process of second mode and associated viscous dissipation have been addressed in our recent work \cite{Zhang2015}.   In our study, the wall temperature is about 300~K, so $\mu^\prime$ was chosen to be $\mu^\prime=2.0\mu$, its value in low pressure nitrogen at $T = 293$~K.

\begin{figure}
\includegraphics[width=9 cm]{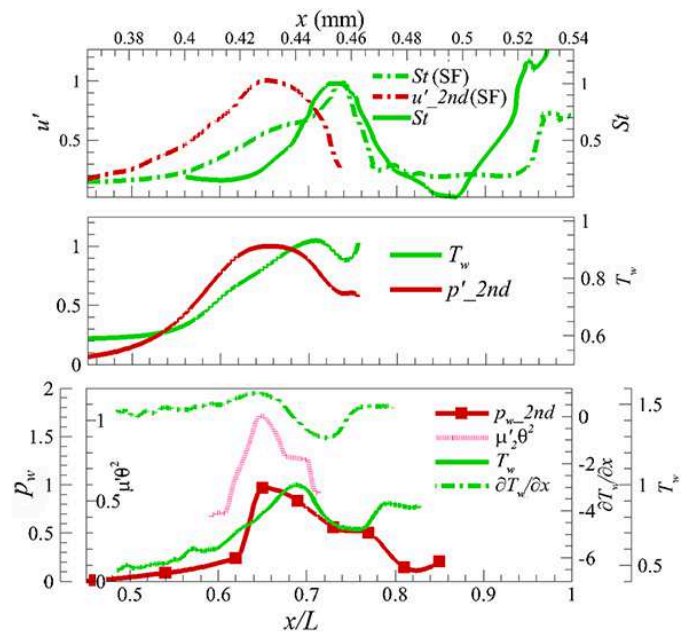}
\caption{\label{fig:comp} Comparison of streamwise evolution of instability  amplitudes with spanwise-averaged surface temperature. Top:\ DNS; middle:\ PSE;  bottom:\  experimental results. SF: DNS results from \citet{Sivasubramanian2015}. Dilatational dissipation and streamwise temperature gradient are also plotted in the bottom part.}
\end{figure}

Evolution of the viscous dissipation due to dilatation can be calculated from the time-averaged $\vartheta^2$ field, obtained from the PIV measurements. The bottom part of Figure \ \ref{fig:comp} displays more clearly the peak value of $\mu^\prime \vartheta^2$, as well as their correlation with the spatial evolutions of second-mode wave amplitudes.

The most prominent feature of the second-mode is its high frequency of about 330~KHz, which is believed to be one of the most important factors that affect the intensity of the dilatational process and its associated viscous dissipation. For simplicity, we consider the velocity fluctuations in its two-dimensional form

\begin{equation}
(u,v)=(\tilde{u},\tilde{v})e^{i(\omega x/c-\omega t)},
\end{equation}
where $\tilde{u}$ and $\tilde{v}$ denote the fluctuation amplitudes, and  $\omega$ and $c$ denote the angular frequency and wave speed. Thus, the dilatation can be written as

\begin{equation}
\label{eq:theta}
\vartheta=(i\omega\tilde{u}/c+\tilde{v}^\prime)e^{i(\omega x/c-\omega t)}.
\end{equation}
where the prime denotes differentiation with respect to $y$. For a given $\tilde{u}$ and $c$, the higher the frequency is, the more intense the dilatation is. Moreover, the dilatational dissipation for a gas arises from an internal molecular relaxation process, which is, of course, a dispersion of frequency. As interpreted by \citet{Landau1959}, the bulk viscosity that leads to attenuation of dilatation waves increases monotonically with frequency.

Fig.~\ref{fig:comp} presents both experimental and numerical results of the streamwise evolution of  second-mode amplitudes and spanwise-averaged surface temperatures. The trends observed in all three methods are similar.  Extraction of first- and second-mode amplitudes from PCB pressure time series is described in detail in \cite{Zhang2015,Zhu2016}. The second mode increases dramatically starting at $x/L=0.50$, reaches its maximum at about $x/L=0.65$, then decays to nearly zero at $x/L=0.80$.  In contrast, the first mode increases continuously along the streamwise direction. The temperature overshoot at OS1 is directly related to the evolution of second mode and dilatational dissipation, but the peak of OS1 slightly shifts downstream.

Theoretically, we may study the time-averaged surface temperature distribution based on the steady-state energy equation

\begin{equation}
\rho C_v(q\cdot\nabla)T=-p\vartheta +\phi+\nabla\cdot(k\nabla T),
\end{equation}
where $\rho$, $C_v$, $T$, $p$, and $k$ are, respectively, the density, specific heat, temperature, pressure, and thermal conductivity of the gas, and $\phi$ denotes the time-average of $\Phi$. For 2-D flow near the wall, the equation can be simplified to

\begin{equation}
\label{eq:2D}
\rho C_v u\frac{\partial T}{\partial x} = k\frac{\partial^2T}{\partial y^2} + \phi(x,y),
\end{equation}
where the three terms represent, respectively, heat convection in $x$, conduction in $y$, and heat sources caused by dissipation.  In order to be able to integrate (\ref{eq:2D}), we linearize this equation by replacing $u(y)$ by its mean value $U$, and assuming $C_v$ and $k$ to be constants. The boundary conditions at an adiabatic wall and infinity are respectively

\begin{equation}
\frac{\partial T}{\partial y} = 0 \text{ at } y = 0; \ \ \ T = 0 \text{ at } y=\infty.
\end{equation}
The initial temperature profile at $x$=0 is assumed to be $T_0(y)$.
The linearized problem can be solved using Laplace transformation, which yields the wall temperature $T_w = T(x,0)$:

\begin{eqnarray}
& T_w & = T_1+T_2; \\
& T_1 &=\frac{2}{\sqrt{\pi}}\int_0^\infty T_0\left(2h\sqrt{\frac{kx}{C_vU}}\right)\mathrm{e}^{-h^2}\mathrm{d}h; \\
& T_2 &=\frac{1}{\sqrt{\pi}}\int^x_0\int_0^\infty \phi\left(\tau,2h\sqrt{\frac{k(x-\tau)}{C_vU}}\right)\mathrm{e}^{-h^2}\mathrm{d}h\mathrm{d}\tau.
\end{eqnarray}

Neglecting the variation of the heat source in the $y$ direction, $T_2$ can be further simplified to

\begin{eqnarray}
& T_2 & =\int^x_0 \phi(\tau)\mathrm{d}\tau.
\end{eqnarray}
While $T_1$ represents the convection--conduction process without heat sources, $T_2$ stands for the integration of heat sources along the streamwise direction. The $x$-wise gradient of $T_w$ is given by that of $T_1$ and $T_2$:
\begin{eqnarray}
\frac{\partial T_1}{\partial x}& = & \frac{4}{\sqrt{\pi}}\int_0^\infty \frac{\mathrm{d}T_0}{\mathrm{d}y}h\sqrt{\frac{k}{xC_vU}}\mathrm{e}^{-h^2}\mathrm{d}h;\\
\frac{\partial T_2}{\partial x}& = &\phi(x) \label{eq:gradient}.
\end{eqnarray}

The sign of $\frac{\partial T_1}{\partial x}$ depends on the initial distribution of $T_0(y)$. For hypersonic boundary layer, $\frac{\mathrm{d}T_0}{\mathrm{d}y} < 0$ and $\frac{\mathrm{d}^2T_0}{\mathrm{d}y^2}<0$, thus $\frac{\partial T_1}{\partial x}<0$, which has the effect of cooling the wall, but its magnitude decreases with $x$. When $\phi$ reaches its maximum value, it begins to decay although $T_2$ still increases. Thus, if $\phi(x)$ decreases so rapidly that

\begin{eqnarray}
\phi(x) \leq \left| \frac{\partial T_1}{\partial x} \right|
\end{eqnarray}
a peak value of $T$ occurs, which slightly shifts downstream  relative to the $\Phi$-peak, as observed in Fig.~\ref{fig:comp}.

\begin{figure}
\includegraphics[width=9 cm]{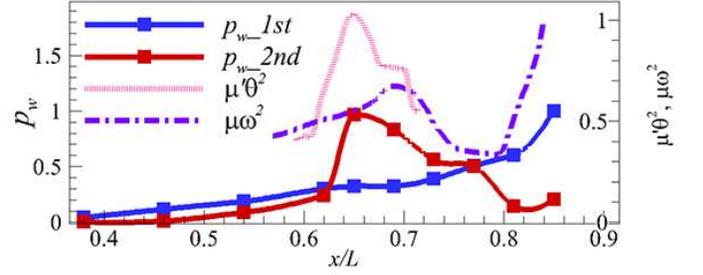}
\caption{\label{fig:both} Comparison of streamwise evolution of instability amplitudes and viscous dissipation caused by shear and dilatation, respectively. Obtained from time-averaged PCB and PIV results.}
\end{figure}

To verify Eq.~(\ref{eq:gradient}), we compare the streamwise surface temperature gradient and viscous dissipation functions, which are obtained, respectively, from TSP and PIV measurements. As shown in the bottom part of Fig.~\ref{fig:comp}, the peak positions of surface temperature gradient and viscous dissipation agree very well  at $x/L$=0.65, which justifies  a posteriori the approximation to linearize Eq.~(\ref{eq:2D}).

The transition scenario proposed by our previous  investigations \cite{Zhang2015,Zhu2016} indicates that although the second mode enjoys the most rapid growth, it grows, saturates, and attenuates along the streamwise direction, accompanied with a high-value dilatational viscous dissipation.  This leads to a rapid temperature overshoot at OS1.  Whereas the first mode grows continuously, benefitting from the phase-locked interaction with the second mode, and finally triggers the transition.   Coupled with the evolution of multi-mode instabilities, different forms of viscous dissipation come into play along the streamwise direction, as shown in Fig.~\ref{fig:both}. The growth of the first-mode instability increases the shear-induced viscous dissipation,  thus leading to a second temperature overshoot at OS2. The flow visualization in Fig.~\ref{fig:tv} also shows that the boundary layer becomes  ``thinner" at $x/L$=0.76, indicating a growth of wall-shear stress in this region. The proposed physical mechanism at OS2 is also supported by the DNS results of \citet{Franko2013}, which shows that the interaction of the first-mode oblique waves is most probably leading  to the generation of strong streamwise vorticity and consequent growth of wall-shear stress.

It is remarkable that OS1 due to dilatational instability is even stronger than OS2 due to shear instability, which might brings in a higher risk of breaking the TPS tile. According to the report of the Columbia accident investigation board, a local damage on the TPS tile might lead to a disastrous accident \cite{Report2003}.   When considering hypersonic flight safety, attention should be paid to the dilatational heating mechanism identified herein.

In summary,  our results identified a new mechanism  to generate  local overshoot in temperature and  rise  in turbulence production.    The rapid growth of second-mode instabilities, accompanied with severe high-frequency compression and expansion of fluid, produces significant aerodynamic heating in a small region.  The fluid's temerature increases when the gas passes through that region, resulting in a sudden increase of surface temperature.   When the dilatation process is weakened downstream, the temperature may decrease due to heat conduction in the  $y$-direction. Finally, the surface temperature regrows again due to the increase of first-mode waves, which ultimately triggers  boundary layer transition to turbulence.

Aerodynamic heating during transition has also been examined by solving the steady-state energy equation. It is found that the rate of growth and decay of second mode has strong effect on the temperature distribution.

The discovered phenomenon and its physical mechanism is of value in developing new engineering models to predict hypersonic aerodynamic heating and corresponding TPS for hypersonic vehicles. More attention should be paid to the TPS design against the local impact of dilatation-induced aerodynamic heating.

\begin{acknowledgments}
This work was supported by the National Natural Science Foundation of China under grant numbers 109103010062 and 10921202.
\end{acknowledgments}

%


\end{document}